\documentclass[prd,amsmath,amssymb,superscriptaddress,nofootinbib,twocolumn]{revtex4}

\usepackage{amsfonts}
\usepackage{multirow}
\usepackage{makecell}
\usepackage{mathrsfs}
\usepackage{graphicx}
\usepackage{amsmath}
\usepackage{amssymb}
\usepackage{bm}
\usepackage{bbm}
\usepackage{xcolor}
\usepackage{ulem}
\usepackage{array}
\usepackage{diagbox}
\usepackage{float}
\usepackage{subfigure}
\usepackage[
    colorlinks,
    linkcolor=blue,
    anchorcolor=black,
    citecolor=blue
]{hyperref}

\newcommand{\nn}{\nonumber}

\newcommand{\beq}{\begin{equation}}
\newcommand{\eeq}{\end{equation}}
\newcommand{\bqa}{\begin{eqnarray}}
\newcommand{\eqa}{\end{eqnarray}}

\newcommand{\bseq}{\begin{subequations}}
\newcommand{\eseq}{\end{subequations}}

\makeatletter

\begin{document}

\title{Novel azimuthal observables from two-photon collision at $e^+e^-$ colliders
}
\author{Yu Jia~\footnote{jiay@ihep.ac.cn}}
\affiliation{Institute of High Energy Physics,
Chinese Academy of Sciences, Beijing 100049, China\vspace{0.2cm}}
\author{Jian Zhou~\footnote{jzhou@sdu.edu.cn}}
\affiliation{Key Laboratory of
Particle Physics and Particle Irradiation (MOE),Institute of
Frontier and Interdisciplinary Science, Shandong University,
(QingDao), Shandong 266237, China \vspace{0.2cm}}
\affiliation{Southern Center for Nuclear-Science Theory (SCNT), Institute of Modern Physics, Chinese Academy of Sciences, HuiZhou, Guangdong
516000, China\vspace{0.2cm}}
\author{Ya-jin Zhou~\footnote{zhouyj@sdu.edu.cn}}
\affiliation{Key Laboratory of
Particle Physics and Particle Irradiation (MOE),Institute of
Frontier and Interdisciplinary Science, Shandong University,
(QingDao), Shandong 266237, China \vspace{0.2cm}}

\date{\today}

\begin{abstract}
In this work we advocate a set of novel azimuthal-angle-related observables associated with exclusive hadron production from
two-photon fusion at $e^+ e^-$  colliders, taking the $\gamma\gamma\to \pi\pi$ as a benchmark process. 
As a direct consequence of the linearly polarized quasi-real photons emitted off the
electron and positron beams, the $\cos 2\phi$ azimuthal asymmetry in dipion production is predicted within 
the transverse-momentum-dependent (TMD) factorization framework.
In numerical analysis, we take the helicity amplitudes of $\gamma\gamma\to \pi\pi$ determined from the partial wave solutions in dispersion relation as input,
and find that the  predicted $\cos2\phi$ azimuthal modulation  may reach 40\% for the typical kinematical setup of {\tt Belle 2} and {\tt BESIII} experiments. 
Future accurate measurement of this azimuthal asymmetry may facilitate
the direct extraction of the relative phase between two helicity amplitudes with photon helicity configurations $++$ and $+-$.
This knowledge provides a valuable input for the dispersive determination of the hadronic light-by-light (Hlbl) contributions.
\end{abstract}

\maketitle

\noindent{\color{blue}\it Introduction.}  Exclusive hadronic production in two-photon collisions at $e^+e^-$ colliders,
in which two quasi-real photons radiated off the electron and positron fly nearly parallel to the beam pipe,
has long been an important research frontier of QCD~\cite{Budnev:1975poe,Two:photon:physics:book}.
Exclusive meson pair production in two-photon fusion with the invariant mass below 2 GeV
plays an indispensable role in revealing the $C$-even resonant structures and advancing our understanding toward the
internal structure of mesons~\cite{CrystalBall:1985mzc,TPCTwoGamma:1986tjf,CELLO:1992iai,Belle:2007ebm,Belle:2009ylx,Belle:2004bpk,Belle:2003xlt,Belle:2006whh}.

In the standard treatment of meson pair production from two-photon collision at $e^+e^-$ colliders,
one usually invokes the equivalent photon approximation (EPA) and QED collinear factorization,
so that the production cross section can be expressed as the product of the photon parton distribution functions (PDFs) inside an
electron and positron folded with the partonic cross section for $\gamma\gamma \to M \overline{M}$~\cite{Budnev:1975poe,Two:photon:physics:book}.
Classical observables in QED collinear factorization are the invariant mass and rapidity distributions of the meson pair.
Nevertheless, in a typical two-photon collision in the realistic $e^+e^-$ experiment, the $M$ and $\overline{M}$ always fly {\it nearly}, but not {\it exactly},
back-to-back in the transverse plane. In the {\it correlation} limit where the total transverse momentum of the meson pair is much smaller than the transverse momentum carried by each
individual meson, the imbalance of the mesonic transverse momenta implies that the transverse momenta of the incident photons
should not be neglected. It has been recognized  that the quasi-real photon emitted from a charged particle
is strongly linearly polarized, with the polarization vector aligned with
its transverse momentum direction (see Ref.~\cite{Pisano:2013cya} for example).
This fact can also been readily seen from the classical electrodynamics \cite{Jackson}.
The main objective of this work is to advocate a class of novel azimuthal-angle-dependent observables
in $e^+e^-$ experiments, which is a direct consequence of the strongly linearly-polarized photon.
Somewhat surprisingly, this new, uncharted territory of two-photon physics at $e^+e^-$ colliders,
seems to have been completely overlooked in the preceding work~\footnote{It is worth
mentioning that, in the past few years the two-photon physics program has witnessed a renaissance in the ultraperipheral collisions (UPCs)
in relativistic heavy ion experiments. Since the coherent photons radiated off the heavy gold/lead nucleus are highly linearly polarized,
a sizable $\cos 4\phi$ azimuthal asymmetry is anticipated in the Breit-Wheeler process $\gamma\gamma\to l^+l^-$~\cite{Li:2019yzy,Li:2019sin}.
This prediction has soon been confirmed by {\tt STAR} experiment~\cite{STAR:2019wlg}, and
a flurry of theoretical efforts have been dedicated to utilize the two-photon programs at UPCs to 
study nuclear structure~\cite{Xiao:2020ddm,Xing:2020hwh,Zha:2020cst,Brandenburg:2021lnj,Hagiwara:2020juc,Hagiwara:2021xkf,
Lin:2022flv,Wang:2022gkd,Zhou:2022gbh,STAR:2022wfe,Zhao:2023nbl,Li:2023yjt,Xie:2023vfi,Shi:2023nko,Shi:2024gex,Zhang:2024mql,Lin:2024mnj,Yu:2024icm,Mantysaari:2023prg,Linek:2023kga, Taels:2022tza, Hauksson:2024bvv, Eskola:2022vpi},
exotic hadron states~\cite{Niu:2022cug}, and new physics via novel polarization-dependent observables in UPCs~\cite{Xu:2022qme,Shao:2023bga}.}.

Among various meson pair production channels, $\gamma \gamma \rightarrow \pi \pi$ constitutes the cleanest and most important one.
The precise knowledge about this channel not only provides valuable information of the scalar resonance $\sigma$, $f_0(980)$ as well as the tensor
resonance $f_2(1270)$, but serves the important measure for the electromagnetic polarizabilities of pions~\cite{Hoferichter:2011wk,Dai:2016ytz}.
More importantly, the azimuthal asymmetry, which can be {\it directly} accessed experimentally,
encodes the message of the relative phase between two helicity amplitudes with photon helicity configurations $++$ and $+-$.
This is in sharp contrast with all preceding studies, where the phases of the helicity amplitudes are extracted
through some {\it indirect} method, {\it e.g.}, by combining dispersive technique and experimental input.
The accurate knowledge of the corresponding partial wave amplitudes provides the key input
for the dispersive determination of the hadronic light-by-light contribution (Hlbl),
which comprises a major source of uncertainty in theoretical prediction of
muon anomalous magnetic moment~\cite{Pascalutsa:2010sj,Dai:2017cvz,Colangelo:2017qdm,Danilkin:2021icn}.

In this work, we take $\gamma\gamma\to \pi\pi$ as a benchmark process to showcase the new azimuthal observables
originating from the photon linear polarization.  Concretely speaking, we urge our experimental colleagues
at {\tt Belle 2} and {\tt BESIII} to measure the dipion azimuthal asymmetries from the two-photon fusion in $e^+e^-$ collisions.
In contrast, a large portion of  $\pi^+ \pi^-$ events at UPCs in heavy ion experiments would come from $\rho^0$ decay via photon-pomeron fusion,
which is difficult to be accurately accounted in a model-independent way.
Therefore, as far as the dipion azimuthal asymmetry is concerned, benefiting from the very high luminosity,
the $e^+e^-$ colliders exemplified by {\tt Belle 2} and {\tt BESIII} experiments,
appear to be the cleaner and superior playground than UPCs.

\vspace{0.2cm}

\noindent{\color{blue}\it \texorpdfstring{$\pi^+\pi^-$}{} production from \texorpdfstring{$e^+e^-$}{} collisions in TMD factorization.}
Let us specialize to the following photon-induced reaction:
\beq
e^+(l_1)+e^-(l_2) \rightarrow e^+(l_1')+e^-(l_2')+ \pi^+(p_1)+\pi^-(p_2),
\eeq
where the symbols in parentheses label the four momenta of the corresponding particles, and the momenta carried by two radiated quasi-real photons are $k_1$ and $k_2$, subject to the momentum conservation
$k_1+k_2=p_1+p_2$. The square of the invariant mass of the $\pi^+\pi^-$ pair is $Q^2\equiv (p_1+p_2)^2$.
It is convenient to introduce two transverse momenta ${\bf P}_\perp \equiv \frac{{\bf p}_{1\perp}-{\bf p}_{2\perp}}{2}$, and
${\bf q}_\perp\equiv {\bf p}_{1\perp}+{\bf p}_{2\perp}={\bf k}_{1\perp}+{\bf k}_{2\perp}$.
The azimuthal angle is defined by $\cos\phi \equiv \hat{\bf P}_\perp \cdot \hat{\bf q}_\perp$.
In the correlation limit $\vert {\bf q}_\perp\vert \ll \vert {\bf P}_\perp\vert$,
one can approximate ${\bf P}_\perp\approx {\bf p}_{1\perp} \approx -{\bf p}_{2\perp }$,
since $\pi^+$ and $\pi^-$ fly nearly back to back in the transverse plane.

As mentioned before, the quasi-real photons emitted from unpolarized electrons and positrons are linearly polarized,
with the polarization vectors aligned with their transverse momenta.
As a consequence, to access the azimuthal dependent observables in the correlation limit, one ought to utilize
the transverse-momentum-dependent (TMD) factorization formalism, rather than the standard collinear factorization approach widely used in the preceding studies
in two-photon physics. In TMD factorization, the azimuthal-dependent cross section can be expressed as the convolution of
the short-distance part and photon TMD parton distributions of the electron and positron.

In analogy to the operator definition of the gluon TMDs in QCD~\cite{Mulders:2000sh},
the photon TMD PDFs in QED are defined by~\cite{Pisano:2013cya}
\bqa \nn
&& \int \frac{dy^- d^2y_\perp}{P^+(2\pi)^3} e^{ik \cdot y} \langle e |
F_{+}^\mu(0)F_{+}^\nu(y)  |e
\rangle \big|_{y^+=0}
\label{photon:TMD:distributions:def}
\\
&& = \frac{\delta_\perp^{\mu \nu}}{2}
xf(x,k_\perp^2)+ \left (\frac{k_\perp^\mu
k_\perp^\nu}{k_\perp^2}-\frac{\delta_\perp^{\mu\nu} }{2}\right )
 xh_1^{\perp }(x,k_\perp^2),
\eqa
where $f$ and $h_1^{\perp}$ signify the
unpolarized and linearly-polarized photon TMD distributions, respectively. $P^+$ is the longitudinal momentum of the electron, and $x$ is the longitudinal momentum fraction of the electron carried by the photon, where we have used light-cone coordinates of a vector $v^\mu = (v^+, v^-, {\bm v}_\perp)$ with $v^\pm = (v^0 \pm v^3)/\sqrt{2}$. $k_\perp$ is the transverse momentum of the photon. $F_{+}^\mu(y)$ is the electromagnetic field tensor with $y$ being the space-time position. The transverse metric tensor in \eqref{photon:TMD:distributions:def} is defined by
$\delta_\perp^{\mu\nu}=-g^{\mu\nu}+ \frac{p^\mu n^\nu+p^\nu n^\mu}{ p\cdot n}$ with
$n^\mu=(1,-1,0,0)/ \sqrt{2}$, and $k_\perp^2=\delta_\perp^{\mu\nu} k_{\perp\mu} k_{\perp\nu}$.

In contrast to the photon TMD PDFs of a large nucleus in UPCs, the photon TMDs of an electron or positron can be rigorously accounted in perturbation theory.
At the lowest order in QED coupling, one has
\bseq
\bqa
f(x,k_\perp^2)&=& \frac{\alpha_e}{2\pi^2} \frac{1+(1-x)^2}{x} \frac{k_\perp^2}{(k_\perp^2+ x^2 m_e^2)^2},
\\
h_1^{\perp}(x,k_\perp^2)&=& \frac{\alpha_e}{\pi^2} \frac{1-x}{x} \frac{k_\perp^2}{(k_\perp^2+ x^2 m_e^2)^2},
\eqa
\eseq
with $m_e$ signifying the electron mass.
Note that the photon TMD PDFs do not acquire scale dependence due to the absence of initial and final-state radiation.
In passing we remark that the degree of linear polarization of photon increases as $x$ decreases,
similar to the QCD case~\cite{Metz:2011wb}.

For latter use, let us specify the polarization vectors of the first photon with definite helicities:
\beq
\epsilon^\mu(k_1,\pm) = \frac{1}{\sqrt{2}}(0,\mp 1,-i,0).
\label{pvector:1:plus:minus}
\eeq
The polarization vector of the second photon is defined to be
$\epsilon^\mu(k_2,\pm)= \epsilon^\mu(k_1,\mp)$, following Jacob-Wick's second particle
phase convention~\cite{Haber:1994pe}.

It is constructive to reexpress the rank-2 tensors in \eqref{photon:TMD:distributions:def}
in terms of the photon's polarization vectors:
\bseq
\bqa
& & \delta_\perp^{\mu \nu} = \epsilon^\mu(k_j,+) \epsilon^{*\nu}(k_j,+)+\epsilon^\mu(k_j,-) \epsilon^{*\nu}(k_j,-),
\\ \nn
&& \delta_\perp^{\mu\nu}- 2\frac{k_{j\perp}^\mu
k_{j\perp}^\nu}{k_{j\perp}^2}
  = e^{\pm (-1)^j 2i\phi_j} \epsilon^\mu(k_j,\pm) \epsilon^{*\nu}(k_j,\mp), \\
\eqa
\eseq
with $j=1,2$. For definiteness, we have chosen ${\bf P}_\perp$ to align with the $x$-axis, and
$\phi_j$ represents the azimuthal angle between ${\bf k}_{j\perp}$ and ${\bf P}_\perp$.

After some straightforward manipulation, we derive the semi-inclusive dipion cross section for the reaction
$e^+ e^-\rightarrow e^+  e^- + \pi^+ \pi^-$ in TMD factorization framework~\footnote{In contrast to dipion production from two photon fusion in UPCs~\cite{Klusek-Gawenda:2013rtu},
a simplifying feature in $e^+e^-$ experiment is that \eqref{semi-inclusive:production:cross:section} does not involve integration over the impact parameter,
since electron and positron are structureless point-like particles.}:
\bqa
& & \frac{d\sigma}{d^2 {\bf p}_{1\perp} d^2 {\bf p}_{2\perp} dy_1 dy_2 } =
\frac{1}{16 \pi^2 Q^4} \int d^2 {\bf k}_{1\perp} d^2 {\bf k}_{2\perp}
\nn\\
& \times & \delta^2( {\bf q}_\perp-{\bf k}_{1\perp}-{\bf k}_{2\perp})  x_1 x_2
\nn\\
& \times &  \bigg \{ \frac{1}{2}\left( |M_{+-}|^2 +|M_{++}|^2 \right)\,f(x_1,k_{1\perp}^2)f(x_2,k_{2\perp}^2)
\nn \\
&-&
\cos (2\phi_1) {\rm Re} [M_{++} M_{+-}^*] \, f(x_2,k_{2\perp}^2) h_1^{\perp }(x_1,k_{1\perp}^2)
\nn\\
&-& \cos (2\phi_2) {\rm Re}[M_{++} M_{+-}^*] \,  f(x_1,k_{1\perp}^2) h_1^{\perp }(x_2,k_{2\perp}^2)
\nn\\
&+&  \frac{1}{2} \Big[ \cos 2(\phi_1-\phi_2) |M_{++}|^2+\cos 2(\phi_1+\phi_2) |M_{+-}|^2  \Big]
\nn\\
&\times &  h_1^{\perp }(x_1,k_{1\perp}^2) h_1^{\perp }(x_2,k_{2\perp}^2)  \bigg \},
\label{semi-inclusive:production:cross:section}
\eqa
where $y_1$ and $y_2$ denote the rapidities of $\pi^+$ and $\pi^-$,
and $M_{\lambda_1, \lambda_2}(Q,\theta, \phi_i)$ signifies the helicity amplitude for
$\gamma(k_1,\lambda_1)\gamma(k_2,\lambda_2)\to \pi^+(p_1)\pi^-(p_2)$.
The photon's longitudinal momentum fractions $x_{1,2}$ are constrained by the conditions
$x_1= \sqrt{\frac{{\bf P}_\perp^2+ m_\pi^2}{s}}(e^{y_1}+e^{y_2})$, $x_2= \sqrt{\frac{{\bf P}_\perp^2+m_\pi^2}{s}}(e^{-y_1}+e^{-y_2})$,
respectively.
To arrive at the above result, we have employed the parity conservation condition $M_{\lambda_1, \lambda_2} = M_{-\lambda_1, -\lambda_2}$.

Equation~\eqref{semi-inclusive:production:cross:section} constitutes the key formula of this work.
Note this cross section is differential with respect to the transverse momenta of $\pi^+$ and $\pi^-$.
As anticipated, only the first term in the right-handed side of \eqref{semi-inclusive:production:cross:section},
proportional to $|M_{++}|^2+|M_{+-}|^2$ convoluted with the unpolarized photon PDF $f(x,k_\perp^2)$, contribute to the integrated cross section.
After integrating over photons' transverse momenta, this term exactly reproduces the prediction made within collinear factorization.
Interestingly, the novel message is conveyed by the second and third terms in \eqref{semi-inclusive:production:cross:section}.
Note that the $\cos 2\phi_i$ terms are accompanied with the
interference between two distinct helicity amplitudes ${\rm Re}[M_{++} M_{+-}^*]$, in convolution with the 
product of the unpolarized photon TMD PDF $f(x_i, k^2_{i\perp})$
and the linearly-polarized photon
TMD PDF $h_1^\perp(x_i,k^2_{i\perp})$.
After integrating over $k_{1\perp}$ and $k_{2\perp}$, the azimuthal dependence of $\cos 2\phi_{1,2}$ conspire into a
non-vanishing $\cos 2\phi$ modulation.
The $\cos 4\phi$ azimuthal modulation arises in a similar manner, which stems form the contribution of the third term.
It is clear that the linear polarization of photons play a crucial role in generating these azimuthal asymmetries.

\vspace{0.2cm}

\noindent{\color{blue}\it Input of helicity amplitudes $M_{++}$ and $M_{+-}$.} It is evident from \eqref{semi-inclusive:production:cross:section}
that future measurements of the $\cos 2\phi$ azimuthal asymmetry provides a powerful means to extracting the relative phase between $M_{++}$ and $M_{+-}$.
The relative magnitude and phase between two helicity amplitudes are of great theoretical interest,
especially in unravelling the resonant structure. For instance, in the latest {\tt PDG} compilation~\cite{PDG:2022pth}, the ratio of $|M_{++}/M_{+-}|$ 
in $\gamma\gamma\to f_2(1270)\to \pi\pi$ has been estimated to be $3.7\pm 0.3^{+15.9}_{-2.9}$.

On the theoretical side, the reactions $\gamma \gamma \rightarrow \pi\pi$ can be tackled by different approaches.
These reactions have been thoroughly investigated within the framework of chiral perturbation theory ($\chi$PT),
with one-loop and two-loop corrections available long ago~\cite{Bijnens:1987dc,Donoghue:1988eea,Oller:1997yg,Burgi:1996qi,Gasser:2005ud}.
One can readily deduce the analytic expressions of $M_{++}$ and $M_{+-}$, with the one-loop accuracy results provided in the supplemental material.

Unfortunately, the $\chi$PT prediction is expected to make reliable prediction only within a rather limited kinematic window, {\it i.e.},
near the dipion threshold region, say, $Q<500$ MeV. To exploit a great amount of data accumulated far above the dipion threshold,
one has to resort to other theoretical approaches.
As the invariant mass increases, the $\pi\pi$ interaction strength increases, and it becomes compulsory to incorporate the
final-state interaction of $\pi\pi$ in a nonperturbative manner in order to give a reliable account of the $\gamma\gamma\to \pi\pi$ amplitude.
In this work we resort to the parametrized forms of the $M_{++}$ and $M_{+-}$ determined through the 
data-driven dispersive approach by Dai and Pennington~\cite{Dai:2014zta},
which is valid over a wide range of invariant mass.
 The dispersive approach has been
widely applied in the studies of exclusive meson production in $\gamma\gamma$ 
fusion~\cite{Mao:2009cc,Dai:2016ytz,Dai:2017cvz,Yao:2020bxx,Hoferichter:2019nlq,Hoferichter:2024fsj}.

Following the convention of \cite{Dai:2014zta}, we expand the helicity amplitudes in terms of
different partial waves:
\bseq
 \bqa
 M_{++}(Q, \theta, \phi)&=&e^2 \sqrt{16\pi} \sum_{J \ge 0}F_{J0}(Q) Y_{J0}(\theta, \phi),
 \\
 M_{+-}(Q, \theta, \phi)&=&e^2 \sqrt{16\pi} \sum_{J \ge  2}F_{J2}(Q) Y_{J2}(\theta, \phi),
\eqa
\eseq
where  $\theta$ and $\phi$ denote the polar and azimuthal angles of the outgoing pions, and  $Y_{Jm}$ signifies the
spherical harmonics.
All the nontrivial dynamics is encapsulated in the reduced matrix elements $F_{J0}(Q)$  and $F_{J2}(Q)$,
which have been determined over a wide kinematic window through global fitting~\cite{Dai:2014zta}.
With the elastic $\pi\pi$ scattering data as input, Ref.~\cite{Dai:2014zta} is able to a give a quite satisfactory account of
the unpolarized dipion production cross section up to $Q=1.5$ GeV.

\vspace{0.2cm}

\noindent{\color{blue}\it Numerical predictions of azimuthal asymmetry.}
To facilitate the comparison between experiment and theory,
we introduce the averaged azimuthal variables as follows,
\beq
\langle \cos( n\phi) \rangle \equiv \frac{ \int\!\! d \sigma \, \cos n \phi}{\int\!\!  d \sigma},
\eeq
with $n=2,4$. For the purpose of illustration, we take the $e^+e^-$ center-of-mass energy to be $\sqrt{s}=10.58$ GeV
at {\tt Belle 2} and $\sqrt{s}=3.77$ GeV
at {\tt BESIII}. We impose the rapidity cut $|y_{1,2}| \le 0.38$ to match the angular coverage of the {\tt Belle 2} and {\tt BESIII} detectors,
{\it e.g.}, $|\cos \theta|<0.6$. In conformity with the correlation limit, we require $\vert {\bf P}_\perp\vert$ to be greater than 100 MeV,
while $\vert {\bf q}_\perp\vert$ is integrated from 0 to 50 MeV.

\begin{widetext}
\vfill\hfill

\begin{figure}[htbp]
\subfigure[]{\includegraphics[angle=0,scale=0.535]{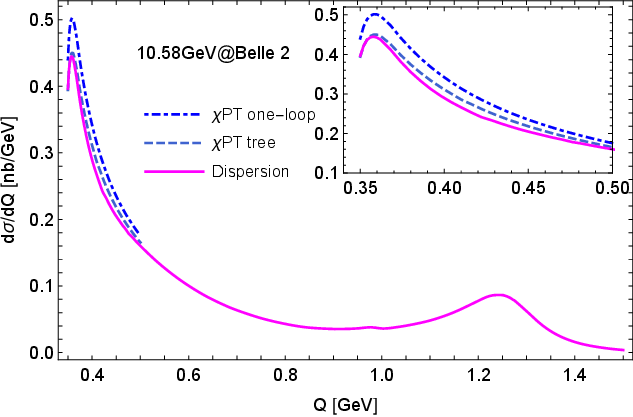}}
\subfigure[]{\includegraphics[angle=0,scale=0.535]{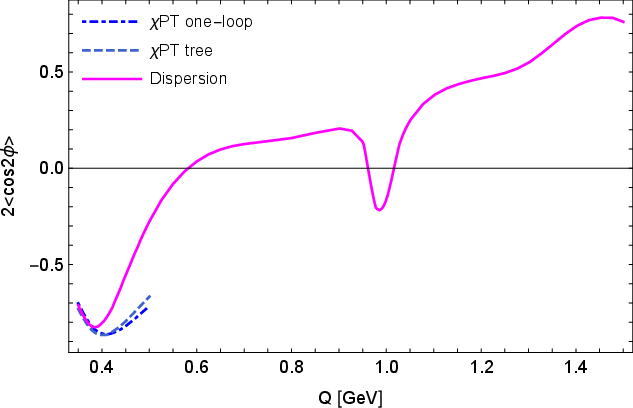}}
\subfigure[]{\includegraphics[angle=0,scale=0.535]{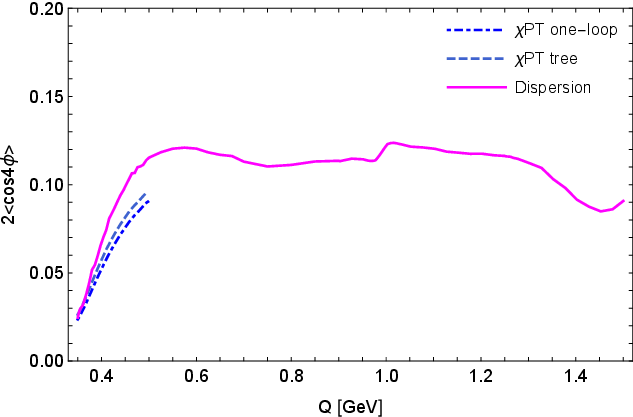}}
\subfigure[]{\includegraphics[angle=0,scale=0.535]{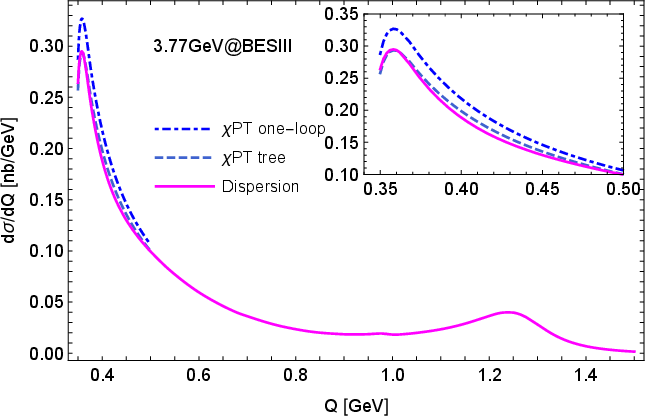}}
\subfigure[]{\includegraphics[angle=0,scale=0.535]{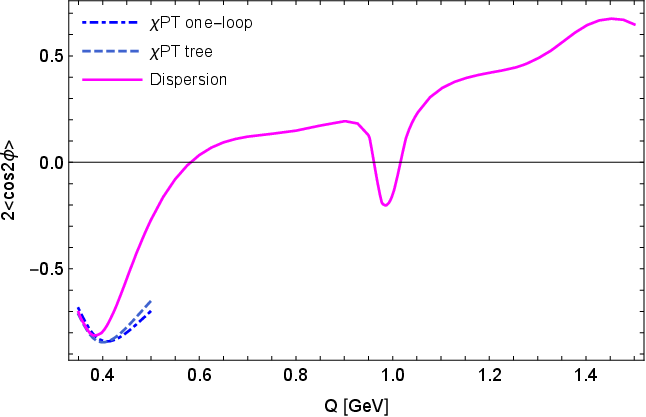}}
\subfigure[]{\includegraphics[angle=0,scale=0.535]{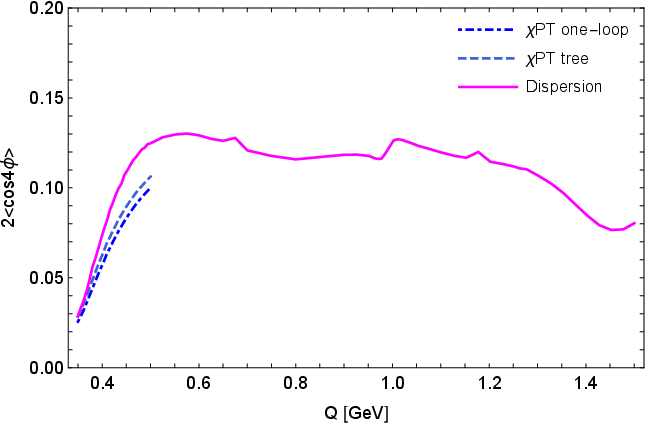}}
\caption{The dipion differential cross section (left panel),
the azimuthal asymmetries $\langle \cos 2\phi \rangle$ (middle panel) and $\langle \cos 4\phi \rangle$ (right panel) as a function of the dipion invariant mass,
for the reaction $e^+e^- \rightarrow \gamma \gamma e^{+}e^{-}\rightarrow \pi^+\pi^- e^{+}e^{-}$ at $\sqrt{s}=10.58$ GeV ({\tt Belle 2}, upper panel) and $\sqrt{s}=3.77$ GeV ({\tt BESIII}, lower pannel).
 The cuts $|y_{1,2}| \le 0.38$ and $\vert {\bf P}_\perp\vert>100$ MeV are imposed, while $q_\perp$ is integrated from 0 to 50 MeV. }
\label{asQ_pi+pi-}
\end{figure}
\end{widetext}

In Fig.~\ref{asQ_pi+pi-} we plot the differential cross section together with the azimuthal asymmetries for $e^+e^- \rightarrow \pi^+\pi^- e^{+}e^{-}$ against the $\pi^+\pi^-$
invariant mass, taking the $e^+e^-$ center-of-mass energy to be 10.58 GeV and 3.77 GeV.
For the sake of comparison, we also present the predictions obtained from the tree-level and one-loop $\chi$PT,
juxtaposed with those obtained from the partial wave solutions provided in Ref.~\cite{Dai:2014zta}.
As expected, at low $\pi^+\pi^-$ invariant mass, the $\chi$PT predictions for the azimuthally-averaged cross sections are in
fair agreement with that obtained from the dispersive analysis.
Curiously, the $\cos 2\phi$ and $\cos 4\phi$ asymmetries predicted by the $\chi$PT predictions start to deviate from those obtained from the
dispersive relation at rather low invariant mass \footnote{It is well-known that ChPT has great difficulty to accurately account the pion-pion $S$-wave phase shift even at rather low energy,
as it does not fully capture the resonant behavior of the $\sigma$ meson \cite{Pelaez:2015qba}.
The pole of this resonance lies around $Q = 440$ MeV, with a width of approximately 200 MeV, which has significant impact on the low-energy $\pi\pi$ scattering.
Therefore, the difference of the predictions between both approaches are understandable.}.

As can be visualized in Fig.~\ref{asQ_pi+pi-}, the $\cos 2\phi$ azimuthal asymmetry is notably pronounced,
and even undergoes a sign change around $Q \approx 0.55$ GeV. Interestingly, a dip-like structure is observed around the resonance $f_0(980)$.
Meanwhile, the $\cos 4\phi$ azimuthal asymmetry increases steadily as the invariant mass rises, which eventually reaches a plateau after $Q>0.6$ GeV,
with the peak asymmetry around 6\%.

Due to the anomalously large amount of $\pi\pi$ events in low invariant mass regime,
{\tt Belle 2} experiment typically chooses a kinematic cut $Q>0.8$ GeV to reduce the background.
With the aforementioned cuts imposed, we predict $\sigma[e^+e^- \rightarrow \pi^+\pi^- e^{+}e^{-}]=0.03$ nb at $\sqrt{s}=10.58$ GeV within
the interval $0.8<Q< 1.5$ GeV. Assuming the integrated luminosity of {\tt Belle} and {\tt Belle 2} until now is about $1500\;{\rm fb}^{-1}$ \cite{Belle:2012iwr, Belle-II:2024vuc, Belle:2024ikp},
one anticipates that there are $4.5\times 10^7$ $\pi^+\pi^-$ events. With such a gigantic number of signal events,
it is reasonable to envisage that the azimuthal asymmetries can be measured to a decent accuracy.

\begin{figure}[htbp]
\centering
\includegraphics[scale=0.6]{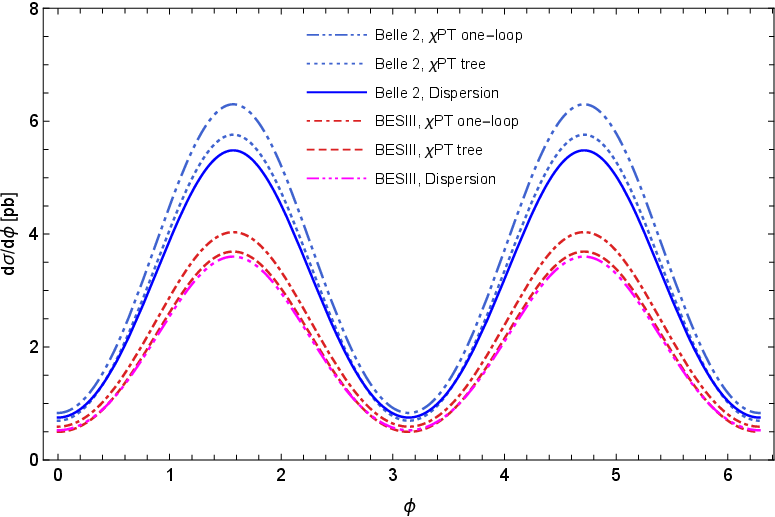}
\includegraphics[scale=0.58]{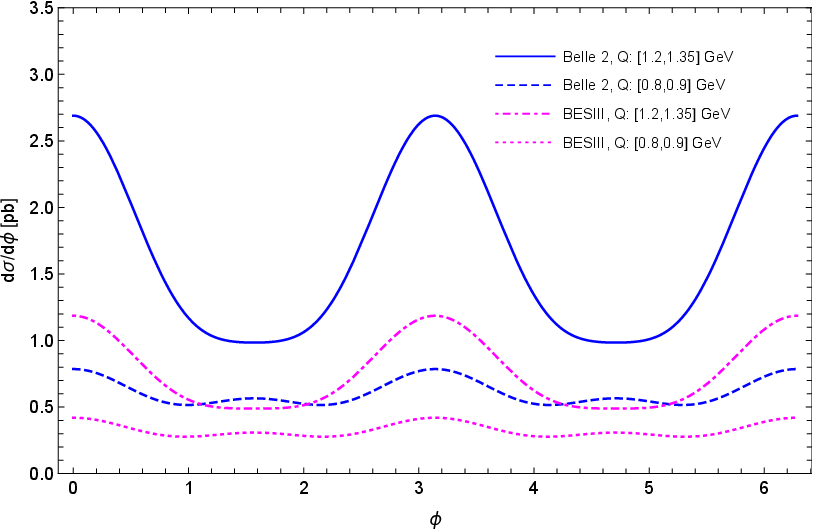}
\caption{The differential cross section of $e^+e^- \rightarrow \gamma \gamma  e^{+}e^{-}\rightarrow \pi^+\pi^- e^{+}e^{-}$ with respect to
the azimuthal angle $\phi$ in various invariant mass intervals. The center-of-mass energy is fixed at $10.58$ GeV and $3.77$ GeV.
The $\pi^+\pi^-$ invariant mass is restricted in the window $0.35<Q< 0.4$ GeV (upper panel), in which the $\chi$PT predictions are compared with
the dispersive predictions.  The lower panel illustrates the cross section differential in $\phi$ predicted from the dispersive approach,
with $\pi^+\pi^-$ invariant mass
restricted in the interval $[0.8,0.9]$ GeV and $[1.2,1.35]$ GeV.}
 \label{asphi_pi+pi-}
\end{figure}

To fathom the azimuthal modulation in a clearer way, we also plot in Fig.~\ref{asphi_pi+pi-} the differential cross section with respect to the azimuthal angle $\phi$.
In the upper panel of Fig.~\ref{asphi_pi+pi-}, we juxtapose the predictions made from the dispersive relations and $\chi$PT in the interval $0.35<Q< 0.4$ GeV.
In such a low $\pi^+\pi^-$ invariant mass, both predictions are quite close to each other.
To emphasize the impact of polarization-dependent observables on probing resonance structures, in the lower panel of Fig.~\ref{asphi_pi+pi-}, we present the predictions within the invariant mass range $1.2 \le Q \le 1.35$ GeV, where the $f_2(1270)$ resonance is prominent, and the range $0.8\le Q \le 0.9$ GeV, where the resonance is absent for comparison. Interestingly, these two differential cross sections possess rather different magnitudes of azimuthal modulation. For the sake of completeness, in the supplemental material we also show azimuthal asymmetries about $\pi^0 \pi^0$ production at both {\tt Belle 2} and {\tt BESIII} energies. We eagerly look forward to the critical test of our predicted azimuthal asymmetries in the {\tt Belle 2} and {\tt BESIII} experiments.

\vspace{0.2 cm}

\noindent{\color{blue}\it Summary.} In this work, we propose novel azimuthal observables in meson pair production from two-photon fusion, focusing on high-luminosity $e^+ e^-$ colliders like {\tt Belle 2} and {\tt BESIII}. For clarity, we use the reaction $\gamma\gamma\to \pi\pi$ as a benchmark to illustrate the azimuthally dependent observables. The key is to employ the fact that the photons emitted from the electron and positron are strongly linearly polarized, with the polarization vectors aligned with their transverse momentum directions.

Employing the TMD factorization, we derive a master formula for the $\pi^+\pi^-$ cross section
differential to the pion's transverse momenta.
In our numerical analysis, we take the helicity amplitudes of $\gamma\gamma\to \pi\pi$ determined from the partial wave solutions in dispersive analysis
as input. Remarkably, adopting the typical kinematic cut at {\tt Belle 2} and {\tt BESIII} experiments, we expect a gigantic number of $\pi^+\pi^-$ signals and predict a pronounced
$\cos 2\phi$ azimuthal asymmetry, which may be as large as 40\%.
This azimuthal asymmetry is sensitive to the relative phase between two helicity amplitudes with photon helicity configurations $++$ and $+-$. Therefore, future
accurate measurement of this type of azimuthal asymmetries is of great phenomenological interest. It can enrich our understanding toward the internal structure of the $C$-even resonances.
It can also offer an important input for the dispersive determination of the Hlbl contributions from two pion intermediate state,
which may help to reduce the theoretical uncertainties in predicting the anomalous magnetic moment of muon.

In our current work, we focus on the $\gamma\gamma \to \pi\pi$ reaction in the nonperturbative regime. It is also interesting to investigate the azimuthal asymmetries with large $\pi\pi$ invariant mass, where perturbative QCD becomes applicable. Future investigations may also include studying azimuthal asymmetries in other channels, such as $\gamma\gamma \to p\bar{p}$, $\gamma\gamma \to \rho\rho$, and $\gamma\gamma \to \gamma\gamma$, both in $e^+e^-$ collisions and UPCs.

\noindent{\color{blue}\it Acknowledgments} We are grateful to Ling-Yun Dai for valuable discussions and comments on the manuscript, 
and for providing us with the numerical data on partial wave solutions of the $\gamma\gamma\to \pi\pi$ helicity amplitudes
compiled in Ref.~\cite{Dai:2014zta}.
We thank Feng-Kun Guo for discussions.
We also thank Sen Jia, Haibo Li, Chengping Shen for discussions on observation prospects of the dipion azimuthal asymmtries
at {\tt BESIII} and {\tt Belle 2} experiments.
The work of Y.~J. is supported in part by the National Science Foundations of China under Grants No.~11925506 and No.~12475090.
The work of J. Z. is supported in part by the National Science Foundations of China under Grant No.~12175118, and No.~12321005.
The work of Y. Z is supported in part by the National Science Foundations of China under Grant No. 12475084 and Shandong Province Natural Science Foundation under Grant No. ZR2024MA012.




\begin{widetext}
\vfill\hfill
\newpage
\appendix

\let\oldaddcontentsline\addcontentsline
\renewcommand{\addcontentsline}[3]{}
\section*{Supplemental material}

\let\addcontentsline\oldaddcontentsline


\subsection{Helicity amplitudes of \texorpdfstring{$\gamma\gamma\rightarrow \pi\pi$}{} in \texorpdfstring{$\chi$}{}PT}
\label{helicity:ampl:chpt}

Near the dipion threshold regime, the $\chi$PT gives a model-independent account of the reaction $\gamma\gamma\to \pi\pi$. 
At one-loop order, the amplitude of $\gamma\gamma\to \pi^+\pi^-$ from $\chi$PT takes the following form~\cite{Bijnens:1987dc,Donoghue:1988eea}:
\beq
{\cal M}(\gamma \gamma \rightarrow \pi^+ \pi^-)=2ie^2 \left [
{\cal C} \epsilon(k_1) \! \cdot \epsilon(k_2)  -\frac{2P_\perp^2}{P_\perp^2+m_\pi^2}
(\epsilon(k_1) \! \cdot \hat P_\perp) (\epsilon(k_2)\cdot \hat P_\perp)
 \right ],
\eeq
with the coefficient $\cal C$ given by
\beq
{\cal C} = 1+\frac{4Q^2}{f^2_\pi}(L_9^r+L_{10}^r)-\frac{1}{16\pi^2 f_\pi^2 }\left ( \frac{3}{2}Q^2 +m_\pi^2 \ln^2 g_\pi(Q^2) +\frac{1}{2} m_K^2 \ln^2 g_K(Q^2)\right ),
\eeq
with the low energy constants $L_9^r+L_{10}^r=1.4\times 10^{-3}$.
The functions $g_\pi$ and $g_K$ are defined by
\beq
 g_\pi(Q^2) =\frac{\sqrt{\frac{Q^2}{m_\pi^2}-4}+\sqrt{\frac{Q^2}{m_\pi^2}}}{\sqrt{\frac{Q^2}{m_\pi^2}-4}-\sqrt{\frac{Q^2}{m_\pi^2}}} \qquad\quad
 g_K(Q^2) =\frac{\sqrt{\frac{Q^2}{m_K^2}-4}+\sqrt{\frac{Q^2}{m_K^2}}}{\sqrt{\frac{Q^2}{m_K^2}-4}-\sqrt{\frac{Q^2}{m_K^2}}}.
\eeq

If chiral loop correction is neglected, {\it e.g.},  if ${\cal C}$ is set to unity, 
one then recovers the tree-level scalar QED prediction by treating $\pi^\pm$ as charged point-like spin-0 particles.

For the production of a pair of neutral pions in two-photon fusion, the one-loop $\chi$Pt prediction reads~\cite{Bijnens:1987dc,Donoghue:1988eea,Oller:1997yg}
\beq
{\cal M}(\gamma \gamma \rightarrow \pi^0 \pi^0) =i{\cal D} 4e^2 \epsilon(k_1) \! \cdot \epsilon(k_2),
\eeq
with
\beq
{\cal D}
 = \frac{
  Q^2}{16 \pi^2 f_\pi^2 } \left [ \left ( 1-\frac{m_\pi^2}{Q^2}  \right )
 \left ( 1+\frac{m_\pi^2}{Q^2} \ln^2 g_\pi(Q^2)  \right ) -\frac{1}{4}  \left ( 1+\frac{m_K^2}{Q^2} \ln^2 g_K(Q^2) \right ) \right ].
\eeq

Inserting the photon polarization vectors given by \eqref{pvector:1:plus:minus} into these amplitudes, one obtains the intended 
helicity amplitudes:
\bseq
 \bqa
{\cal M}^{\pi^+\pi^-}_{++}(Q,\theta,\phi)&=&2  ie^2\frac {
   4 m_{\pi}^2 -(1 - {\cal C})\left (Q^2 - \left (Q^2 - 4 m_{\pi}^2 \right)\cos^2 (\theta) \right) } {Q^2 - \left (Q^2 - 4 m_{\pi}^2 \right)\cos^2 \theta},
\\ 
{\cal M}^{\pi^+\pi^-}_{+-}(Q,\theta,\phi)&=& 2 i e^2 \frac{ (Q^2-4\,m_\pi^2) \sin ^2\theta} {Q^2- (Q^2-4\,m_\pi^2) \cos ^2 \theta} e^{i2\phi},
\\
{\cal M}^{\pi^0\pi^0}_{++}(Q,\theta,\phi)&=&4i  e^2 {\cal D},
\\
{\cal M}^{\pi^0\pi^0}_{+-}(Q,\theta,\phi)&=&0.
\label{chiralamp}
\eqa
\eseq


\subsection{Azimthal asymmetries in \texorpdfstring{$\pi^0 \pi^0$}{}  pair production at {\tt Belle 2} and {\tt BESIII} }

\begin{figure}[htbp]
 \includegraphics[angle=0,scale=0.67]{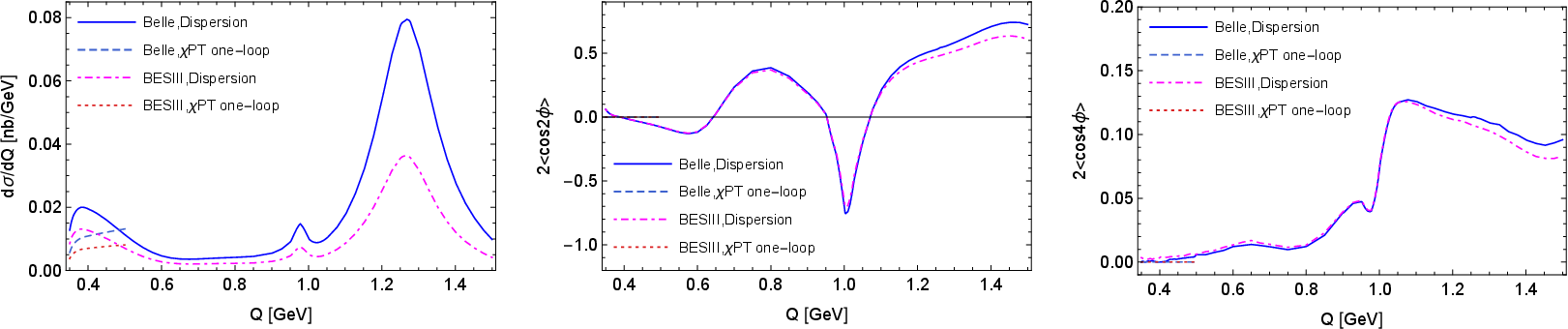}
  \caption{The differential cross section (left panel), 
  $\langle \cos( 2\phi) \rangle$ (middle panel)  and $\langle \cos( 4\phi) \rangle$ (right panel)  
  as  a function of the $\pi^0 \pi^0$ invariant mass, for the reaction $e^+e^- \rightarrow \gamma \gamma e^{+}e^{-}\rightarrow \pi^0\pi^0 e^{+}e^{-}$  
  at $\sqrt{s}= 10.58$ GeV ({\tt Belle 2}) and $\sqrt{s}= 3.77$ GeV ({\tt BESIII}). The rapidity cut $|y_1|, |y_2| \le 0.52$ has been imposed,  while $\vert \bf{q}_\perp \vert$ is integrated from 0 to 50 MeV.
  }
   \label{asQ_pi0pi0}
\end{figure}

In this supplemental material, we present numerical predictions for the azimuthal asymmetries in $\pi^0 \pi^0$ production at {\tt Belle 2} and {\tt BESIII} energies.  
One observes that the azimuthal asymmetries for the $\pi^+ \pi^-$ production at {\tt BESIII} exhibit the similar pattern of invariant mass dependence 
as at {\tt Belle 2}.  However, the asymmetries for $\pi^0\pi^0$ production are drastically different from those for $\pi^+ \pi^-$ production.
It will be interesting to test these predictions in the future measurements at {\tt Belle 2} and {\tt BESIII}.

\end{widetext}


\begin{thebibliography}{99}
\bibitem{Budnev:1975poe}
V.~M.~Budnev, I.~F.~Ginzburg, G.~V.~Meledin and V.~G.~Serbo,
Phys. Rept. \textbf{15}, 181-281 (1975)
doi:10.1016/0370-1573(75)90009-5

\bibitem{Two:photon:physics:book}
H.~Kolanoski,
{\it Two-Photon Physics at $e^+ e^-$ Storage Rings},
Springer Tracts in Modern Physics (STMP, volumn 105) (1984).


\bibitem{CrystalBall:1985mzc}
D.~Antreasyan \textit{et al.} [Crystal Ball],
Phys. Rev. D \textbf{33}, 1847 (1986)
doi:10.1103/PhysRevD.33.1847

\bibitem{TPCTwoGamma:1986tjf}
H.~Aihara \textit{et al.} [TPC/Two Gamma],
Phys. Rev. Lett. \textbf{57}, 404 (1986)
doi:10.1103/PhysRevLett.57.404

\bibitem{CELLO:1992iai}
H.~J.~Behrend \textit{et al.} [CELLO],
Z. Phys. C \textbf{56}, 381-390 (1992)
doi:10.1007/BF01565945

\bibitem{Belle:2007ebm}
T.~Mori \textit{et al.} [Belle],
J. Phys. Soc. Jap. \textbf{76}, 074102 (2007)
doi:10.1143/JPSJ.76.074102
[arXiv:0704.3538 [hep-ex]].

\bibitem{Belle:2009ylx}
S.~Uehara \textit{et al.} [Belle],
Phys. Rev. D \textbf{79}, 052009 (2009)
doi:10.1103/PhysRevD.79.052009
[arXiv:0903.3697 [hep-ex]].

\bibitem{Belle:2004bpk}
H.~Nakazawa \textit{et al.} [Belle],
Phys. Lett. B \textbf{615}, 39-49 (2005)
doi:10.1016/j.physletb.2005.03.067
[arXiv:hep-ex/0412058 [hep-ex]].

\bibitem{Belle:2003xlt}
K.~Abe \textit{et al.} [Belle],
Eur. Phys. J. C \textbf{32}, 323-336 (2003)
doi:10.1140/epjc/s2003-01468-9
[arXiv:hep-ex/0309077 [hep-ex]].

\bibitem{Belle:2006whh}
W.~T.~Chen \textit{et al.} [Belle],
Phys. Lett. B \textbf{651}, 15-21 (2007)
doi:10.1016/j.physletb.2007.05.059
[arXiv:hep-ex/0609042 [hep-ex]].

\bibitem{Pisano:2013cya}
C.~Pisano, D.~Boer, S.~J.~Brodsky, M.~G.~A.~Buffing and P.~J.~Mulders,
JHEP \textbf{10}, 024 (2013)
doi:10.1007/JHEP10(2013)024
[arXiv:1307.3417 [hep-ph]].

\bibitem{Jackson}
J. D. Jackson, Classical Electrodynamics. John Wiley \& Sons, Inc., New York, London, 1962.

\bibitem{Li:2019yzy}
C.~Li, J.~Zhou and Y.~J.~Zhou,
Phys. Lett. B \textbf{795}, 576-580 (2019)
doi:10.1016/j.physletb.2019.07.005
[arXiv:1903.10084 [hep-ph]].

\bibitem{Li:2019sin}
C.~Li, J.~Zhou and Y.~J.~Zhou,
Phys. Rev. D \textbf{101}, no.3, 034015 (2020)
doi:10.1103/PhysRevD.101.034015
[arXiv:1911.00237 [hep-ph]].

\bibitem{STAR:2019wlg}
J.~Adam \textit{et al.} [STAR],
Phys. Rev. Lett. \textbf{127}, no.5, 052302 (2021)
doi:10.1103/PhysRevLett.127.052302
[arXiv:1910.12400 [nucl-ex]].

\bibitem{Xiao:2020ddm}
B.~W.~Xiao, F.~Yuan and J.~Zhou,
Phys. Rev. Lett. \textbf{125}, no.23, 232301 (2020)
doi:10.1103/PhysRevLett.125.232301
[arXiv:2003.06352 [hep-ph]].

\bibitem{Xing:2020hwh}
H.~Xing, C.~Zhang, J.~Zhou and Y.~J.~Zhou,
JHEP \textbf{10}, 064 (2020)
doi:10.1007/JHEP10(2020)064
[arXiv:2006.06206 [hep-ph]].

\bibitem{Zha:2020cst}
W.~Zha, J.~D.~Brandenburg, L.~Ruan, Z.~Tang and Z.~Xu,
Phys. Rev. D \textbf{103}, no.3, 033007 (2021)
doi:10.1103/PhysRevD.103.033007
[arXiv:2006.12099 [hep-ph]].

\bibitem{Brandenburg:2021lnj}
J.~D.~Brandenburg, W.~Zha and Z.~Xu,
Eur. Phys. J. A \textbf{57}, no.10, 299 (2021)
doi:10.1140/epja/s10050-021-00595-5
[arXiv:2103.16623 [hep-ph]].

\bibitem{Hagiwara:2020juc}
Y.~Hagiwara, C.~Zhang, J.~Zhou and Y.~J.~Zhou,
Phys. Rev. D \textbf{103}, no.7, 074013 (2021)
doi:10.1103/PhysRevD.103.074013
[arXiv:2011.13151 [hep-ph]].

\bibitem{Hagiwara:2021xkf}
Y.~Hagiwara, C.~Zhang, J.~Zhou and Y.~j.~Zhou,
Phys. Rev. D \textbf{104}, no.9, 094021 (2021)
doi:10.1103/PhysRevD.104.094021
[arXiv:2106.13466 [hep-ph]].

\bibitem{Lin:2022flv}
S.~Lin, R.~J.~Wang, J.~F.~Wang, H.~J.~Xu, S.~Pu and Q.~Wang,
Phys. Rev. D \textbf{107}, no.5, 054004 (2023)
doi:10.1103/PhysRevD.107.054004
[arXiv:2210.05106 [hep-ph]].

\bibitem{Wang:2022gkd}
R.~j.~Wang, S.~Lin, S.~Pu, Y.~f.~Zhang and Q.~Wang,
Phys. Rev. D \textbf{106}, no.3, 034025 (2022)
doi:10.1103/PhysRevD.106.034025
[arXiv:2204.02761 [hep-ph]].

\bibitem{Zhou:2022gbh}
J.~Zhou [STAR],
EPJ Web Conf. \textbf{259}, 13014 (2022)
doi:10.1051/epjconf/202225913014

\bibitem{STAR:2022wfe}
M.~Abdallah \textit{et al.} [STAR],
Sci. Adv. \textbf{9}, no.1, eabq3903 (2023)
doi:10.1126/sciadv.abq3903
[arXiv:2204.01625 [nucl-ex]].

\bibitem{Zhao:2023nbl}
Q.~Zhao, Y.~X.~Wu, M.~Ababekri, Z.~P.~Li, L.~Tang and J.~X.~Li,
Phys. Rev. D \textbf{107}, no.9, 096013 (2023)
doi:10.1103/PhysRevD.107.096013
[arXiv:2304.04367 [hep-ph]].

\bibitem{Li:2023yjt}
X.~Li, J.~Luo, Z.~Tang, X.~Wu and W.~Zha,
Phys. Lett. B \textbf{847}, 138314 (2023)
doi:10.1016/j.physletb.2023.138314
[arXiv:2307.01549 [hep-ph]].

\bibitem{Xie:2023vfi}
Y.~P.~Xie and V.~P.~Gon\c{c}alves,
Eur. Phys. J. C \textbf{83}, no.6, 528 (2023)
doi:10.1140/epjc/s10052-023-11720-7

\bibitem{Shi:2023nko}
P.~Shi, X.~Bo-Wen, Z.~Jian and Z.~Ya-Jin,
Acta Phys. Sin. \textbf{72}, no.7, 072503 (2023)
doi:10.7498/aps.72.20230074

\bibitem{Shi:2024gex}
Y.~Shi, L.~Chen, S.~Y.~Wei and B.~W.~Xiao,
[arXiv:2406.07634 [hep-ph]].

\bibitem{Zhang:2024mql}
C.~Zhang, L.~M.~Zhang and D.~Y.~Shao,
[arXiv:2406.05618 [hep-ph]].

\bibitem{Lin:2024mnj}
S.~Lin, J.~Y.~Hu, H.~J.~Xu, S.~Pu and Q.~Wang,
[arXiv:2405.16491 [hep-ph]].

\bibitem{Yu:2024icm}
K.~Yu, J.~Peng, S.~Li, K.~Wu, W.~Xie and F.~Sun,
Phys. Rev. C \textbf{109}, no.6, 064907 (2024)
doi:10.1103/PhysRevC.109.064907

\bibitem{Eskola:2022vpi}
K.~J.~Eskola, C.~A.~Flett, V.~Guzey, T.~L\"oyt\"ainen and H.~Paukkunen,
Phys. Rev. C \textbf{106}, no.3, 035202 (2022)
doi:10.1103/PhysRevC.106.035202
[arXiv:2203.11613 [hep-ph]].

\bibitem{Taels:2022tza}
P.~Taels, T.~Altinoluk, G.~Beuf and C.~Marquet,
JHEP \textbf{10}, 184 (2022)
doi:10.1007/JHEP10(2022)184
[arXiv:2204.11650 [hep-ph]].

\bibitem{Linek:2023kga}
B.~Linek, A.~\L{}uszczak, M.~\L{}uszczak, R.~Pasechnik, W.~Sch\"afer and A.~Szczurek,
JHEP \textbf{10}, 179 (2023)
doi:10.1007/JHEP10(2023)179
[arXiv:2308.00457 [hep-ph]].

\bibitem{Mantysaari:2023prg}
H.~M\"antysaari, F.~Salazar, B.~Schenke, C.~Shen and W.~Zhao,
Phys. Rev. C \textbf{109}, no.2, 024908 (2024)
doi:10.1103/PhysRevC.109.024908
[arXiv:2310.15300 [nucl-th]].

\bibitem{Hauksson:2024bvv}
S.~Hauksson, E.~Iancu, A.~H.~Mueller, D.~N.~Triantafyllopoulos and S.~Y.~Wei,
JHEP \textbf{06}, 180 (2024)
doi:10.1007/JHEP06(2024)180
[arXiv:2402.14748 [hep-ph]].


\bibitem{Niu:2022cug}
P.~Y.~Niu, E.~Wang, Q.~Wang and S.~Yang,
[arXiv:2209.01924 [hep-ph]].

\bibitem{Xu:2022qme}
I.~Xu, N.~Lewis, X.~Wang, J.~D.~Brandenburg and L.~Ruan,
[arXiv:2211.02132 [hep-ex]].

\bibitem{Shao:2023bga}
D.~Y.~Shao, B.~Yan, S.~R.~Yuan and C.~Zhang,
[arXiv:2310.14153 [hep-ph]].

\bibitem{Hoferichter:2011wk}
M.~Hoferichter, D.~R.~Phillips and C.~Schat,
Eur. Phys. J. C \textbf{71}, 1743 (2011)
doi:10.1140/epjc/s10052-011-1743-x
[arXiv:1106.4147 [hep-ph]].

\bibitem{Dai:2016ytz}
L.~Y.~Dai and M.~R.~Pennington,
Phys. Rev. D \textbf{94}, no.11, 116021 (2016)
doi:10.1103/PhysRevD.94.116021
[arXiv:1611.04441 [hep-ph]].


\bibitem{Pascalutsa:2010sj}
V.~Pascalutsa and M.~Vanderhaeghen,
Phys. Rev. Lett. \textbf{105}, 201603 (2010)
doi:10.1103/PhysRevLett.105.201603
[arXiv:1008.1088 [hep-ph]].

\bibitem{Dai:2017cvz}
L.~Y.~Dai and M.~R.~Pennington,
Phys. Rev. D \textbf{95}, no.5, 056007 (2017)
doi:10.1103/PhysRevD.95.056007
[arXiv:1701.04460 [hep-ph]].

\bibitem{Colangelo:2017qdm}
G.~Colangelo, M.~Hoferichter, M.~Procura and P.~Stoffer,
Phys. Rev. Lett. \textbf{118}, no.23, 232001 (2017)
doi:10.1103/PhysRevLett.118.232001
[arXiv:1701.06554 [hep-ph]].

\bibitem{Danilkin:2021icn}
I.~Danilkin, M.~Hoferichter and P.~Stoffer,
Phys. Lett. B \textbf{820}, 136502 (2021)
doi:10.1016/j.physletb.2021.136502
[arXiv:2105.01666 [hep-ph]].

\bibitem{Mulders:2000sh}
P.~J.~Mulders and J.~Rodrigues,
Phys. Rev. D \textbf{63}, 094021 (2001)
doi:10.1103/PhysRevD.63.094021
[arXiv:hep-ph/0009343 [hep-ph]].

\bibitem{Metz:2011wb}
A.~Metz and J.~Zhou,
Phys. Rev. D \textbf{84}, 051503(R) (2011)
doi:10.1103/PhysRevD.84.051503
[arXiv:1105.1991 [hep-ph]].

\bibitem{Haber:1994pe}
H.~E.~Haber, SCIPP-93-49, NSF-ITP-94-30
[arXiv:hep-ph/9405376 [hep-ph]].


\bibitem{Klusek-Gawenda:2013rtu}
M.~Klusek-Gawenda and A.~Szczurek,
Phys. Rev. C \textbf{87}, no.5, 054908 (2013)
doi:10.1103/PhysRevC.87.054908
[arXiv:1302.4204 [nucl-th]].


\bibitem{PDG:2022pth}
R.~L.~Workman \textit{et al.} [Particle Data Group],
PTEP \textbf{2022}, 083C01 (2022)
doi:10.1093/ptep/ptac097



\bibitem{Bijnens:1987dc}
J.~Bijnens and F.~Cornet,
Nucl. Phys. B \textbf{296}, 557-568 (1988)
doi:10.1016/0550-3213(88)90032-6

\bibitem{Donoghue:1988eea}
J.~F.~Donoghue, B.~R.~Holstein and Y.~C.~Lin,
Phys. Rev. D \textbf{37}, 2423 (1988)
doi:10.1103/PhysRevD.37.2423

\bibitem{Oller:1997yg}
J.~A.~Oller and E.~Oset,
Nucl. Phys. A \textbf{629}, 739-760 (1998)
doi:10.1016/S0375-9474(97)00649-0
[arXiv:hep-ph/9706487 [hep-ph]].

\bibitem{Burgi:1996qi}
U.~Burgi,
Nucl. Phys. B \textbf{479}, 392-426 (1996)
doi:10.1016/0550-3213(96)00454-3
[arXiv:hep-ph/9602429 [hep-ph]].

\bibitem{Gasser:2005ud}
J.~Gasser, M.~A.~Ivanov and M.~E.~Sainio,
Nucl. Phys. B \textbf{728}, 31-54 (2005)
doi:10.1016/j.nuclphysb.2005.09.010
[arXiv:hep-ph/0506265 [hep-ph]].

\bibitem{Dai:2014zta}
L.~Y.~Dai and M.~R.~Pennington,
Phys. Rev. D \textbf{90}, no.3, 036004 (2014)
doi:10.1103/PhysRevD.90.036004
[arXiv:1404.7524 [hep-ph]].

\bibitem{Mao:2009cc}
Y.~Mao, X.~G.~Wang, O.~Zhang, H.~Q.~Zheng and Z.~Y.~Zhou,
Phys. Rev. D \textbf{79}, 116008 (2009)
doi:10.1103/PhysRevD.79.116008
[arXiv:0904.1445 [hep-ph]].

\bibitem{Yao:2020bxx}
D.~L.~Yao, L.~Y.~Dai, H.~Q.~Zheng and Z.~Y.~Zhou,
Rept. Prog. Phys. \textbf{84}, no.7, 076201 (2021)
doi:10.1088/1361-6633/abfa6f
[arXiv:2009.13495 [hep-ph]].

\bibitem{Hoferichter:2019nlq}
M.~Hoferichter and P.~Stoffer,
JHEP \textbf{07}, 073 (2019)
doi:10.1007/JHEP07(2019)073
[arXiv:1905.13198 [hep-ph]].

\bibitem{Hoferichter:2020lap}
M.~Hoferichter and P.~Stoffer,
JHEP \textbf{05}, 159 (2020)
doi:10.1007/JHEP05(2020)159
[arXiv:2004.06127 [hep-ph]].

\bibitem{Hoferichter:2024fsj}
M.~Hoferichter, P.~Stoffer and M.~Zillinger,
JHEP \textbf{04}, 092 (2024)
doi:10.1007/JHEP04(2024)092
[arXiv:2402.14060 [hep-ph]].

\bibitem{Pelaez:2015qba}
J.~R.~Pelaez,
Phys. Rept. \textbf{658}, 1 (2016)
doi:10.1016/j.physrep.2016.09.001
[arXiv:1510.00653 [hep-ph]].


\bibitem{Belle:2012iwr}
J.~Brodzicka \textit{et al.} [Belle],
PTEP \textbf{2012}, 04D001 (2012)
doi:10.1093/ptep/pts072
[arXiv:1212.5342 [hep-ex]].

\bibitem{Belle-II:2024vuc}
I.~Adachi \textit{et al.} [Belle-II],
Chin. Phys. C \textbf{49}, 013001 (2025)
doi:10.1088/1674-1137/ad806c
[arXiv:2407.00965 [hep-ex]].

\bibitem{Belle:2024ikp}
I.~Adachi \textit{et al.} [Belle and Belle-II],
JHEP \textbf{10}, 045 (2024)
doi:10.1007/JHEP10(2024)045
[arXiv:2406.04642 [hep-ex]].


\end{thebibliography}
\end{document}